\documentclass[useAMS,usenatbib,referee]{mn2e}
\usepackage[utf8x]{inputenc}
\usepackage{amsmath,amssymb,epsf,amsfonts}
\usepackage{natbib}
\usepackage[pdftex]{graphicx}
\usepackage{times}
\usepackage{ulem}
\usepackage{wrapfig}
\usepackage{epsfig}
\usepackage{appendix}
\def\ifnote{\iffalse}

\title{Limits on the GeV Emission from Gamma-Ray Bursts}
\author[Beniamini P., Guetta D., Nakar E.$\&$ Piran T.]{P. Beniamini $^1$\ , D. Guetta $^2$\ , E. Nakar $^3$\& T. Piran $^4$\\
{$^1$ Racah Institute of Physics, The Hebrew University, Jerusalem 91904, Israel; email: paz.beniamini@mail.huji.ac.il}\\
{$^2$ Osservatorio astronomico di Roma, v. Frascati 33, 00040 Monte Porzio Catone, Italy; email: dafne.guetta@oa-roma.inaf.it}\\
{$^3$ Raymond and Beverly Sackler School of Physics $\&$ Astronomy, Tel Aviv University, Tel Aviv 69978, Israel; email: udini@wise.tau.ac.il}\\
{$^4$ Racah Institute of Physics, The Hebrew University, Jerusalem 91904, Israel; email: tsvi@phys.huji.ac.il}}

\begin{document}
\maketitle
\label{firstpage}

\begin{abstract}
The Large Area Telescope (LAT) on board of the Fermi satellite
detected emission above 20 MeV only in a small fraction of the long
gamma-ray bursts (GRBs) detected by the Fermi Gamma-ray Burst
Monitor (GBM) at 8 keV - 40 MeV.  Those bursts that were detected by
the LAT were among the brightest GBM bursts. We examine a sample of
the most luminous GBM bursts with no LAT detection and obtain upper
limits on their high energy  fluence. We find  an average upper
limit of LAT/GBM fluence ratio of 0.13 for GeV fluence during
$T_{90}$ and an average upper limit ratio of 0.45 for GeV fluence
during the first 600 seconds after the trigger. These ratios
strongly constrain various emission models and in particular rule
out SSC models for the prompt emission. In about a third of both LAT
detected and LAT non-detected bursts, we find that the extrapolation
of the MeV range Band spectrum to the GeV range is larger than the
observed GeV fluence (or its upper limit). While this excess is not
highly significant for any specific burst, the overall excess in a
large fraction of the bursts suggests a decline in the high energy
spectral slope in at least some of these bursts. Possibly an
evidence for the long sought after pair creation limit.
\end{abstract}

\section{Introduction}
\label{sec:int} GRBs were first discovered accidentally in the late
sixties by the Vela satellite \citep{Klebesadel(1973)}. The Vela
satellites detected photons at the 150-750 keV range. This led to
several dedicated missions at the sub-MeV energy range. Indeed the
peak of the energy flux was found to be in this energy range. For
that reason, the main observational and theoretical efforts were
made at this range. EGRET, the GeV detector on board the Compton
Gamma Ray Observatory (CGRO), was the first to detect GRBs at
energies above 100MeV (\citealt{Schneid(1992)},
\citealt{Schneid(1995)}). Two particular results of EGRET were
extremely puzzling. The first, was a detection of an 18 GeV photon
1.5 hours after the burst in GRB 940217 \citep{Hurley(1994)}. The
second, was the detection of a late time rising GeV spectral
component in GRB 941017 \citep{Gonzalez(2003)}. For most of the
bursts, though, there was no high energy (GeV) detection and only
upper limits were obtained \citep{Gonzalez(2005)} indicating that
the GeV fluence is at most comparable to the MeV fluence
\citep{Ando(2008)}.

At June 2008, NASA launched the Fermi gamma-ray space telescope
which consists of two instruments. The LAT is a high energy (30MeV
to 300 GeV) detector. It has a larger energy range, a larger
field-of-view (2.4sr) and a shorter dead time (26 $\mu$sec) than its
predecessor EGRET \citep{Atwood(2009)}. The second instrument on
board of Fermi, the GBM, covers the lower energy range of 8 keV to
40 MeV \citep{Meegan(2009)}, operating as a burst monitor. Up to
February 2010, the LAT has detected 10 bursts at $>$100 MeV \citep{LATtable(2010)}
while at the same time the GBM detected over 400 bursts \citep{GBMtable}.
The LAT detected bursts continue
to show those features first seen by EGRET. Namely, a delayed
(\citealt{Abdo a(2009)}, \citealt{Abdo b(2009)},
\citealt{Ackermann(2010)}, \citealt{Ghisellini(2010)}) and prolonged
\citep{Abdo a(2009)} GeV emission as compared with the sub-MeV. The
LAT fluence in all the LAT detected bursts was lower than, or
comparable to, the GBM fluence \citep{Ghisellini(2010)}.

Understanding the properties of the GeV emission is crucial in order
to answer two fundamental questions concerning GRBs: What is the
emitting mechanism of the prompt gamma rays? and what is the origin
of the high energy emission? For example, many authors
(\citealt{Meszaros(1994)}, \citealt{Waxman(1997)},
\citealt{Wei(1998)}, \citealt{Chiang(1999)},
\citealt{Panaitescu(2000)}, \citealt{Zhang(2001)},
\citealt{Sari(2001)}, \citealt{Guetta(2003)},
\citealt{FanPiran2008}, \citealt{Fan+2008}, \citealt{Nakar(2009)},
\citealt{Dermer(2008)}, \citealt{Finke(2008)}) considered
synchrotron Self Compton (SSC) as a mechanism for strong GeV
emission. The ratio between the SSC Fluence and the synchrotron
fluence is of order\footnote{ Note that the Klein-Nishina limit
plays no important role here, since upscattering of $100$ keV
photons to GeV is done in the Thomson regime by electrons with
Lorentz factor of $\sim 100$.} $(\epsilon_e / \epsilon_B)^{1/2}$ \citep{Sari(1996)}
(where $\epsilon_e$ is the fraction of energy in the electrons and
$\epsilon_B$ is the fraction of energy in the magnetic field), and
is therefore expected to be at least of order unity. Therefore
limits on the GeV to MeV fluence, put strong limits on the SSC model
and the conditions within the emitting region. SSC was also
considered as a plausible radiation mechanism of the sub-Mev
emissions (\citealt{PanaitescuMeszaros(2000)},
\citealt{Kumar(2008)}). \cite{Piran Sari(2009)} have shown that this
model is challenged by the lack of GeV emission that is brighter
than the sub-MeV emission. Tighter limits on the GeV fluence will
rule out this model entirely. \cite{Fan(2009)} has discussed the
implications of a lack of strong GeV emission on the internal shocks model. 
He has shown that the model could work if there is a cutoff in the GeV due to 
pair creation and an SSC peak at the far TeV range.
An alternative source of the GeV emission, motivated by the 
fact that it is prolonged, is that GeV emission 
arises from a different emitting zone than the prompt
emission. This idea has been explored extensively in such models as
the external shock afterglow model (\citealt{Kumar(2009)},
\citealt{Kumar(2010)}, \citealt{Ghisellini(2010)}, \citealt{Piran
Nakar(2010)}, \citealt{Gao(2009)}, \citealt{Granot(2003)},
\citealt{Corsi(2010)}) and the multi-zone internal shocks model
(\citealt{Xue(2008)}, \citealt{Aoi(2009)}, \citealt{Zhao(2010)}).

The low LAT detection rate is consistent with a detection rate
derived by assuming a ratio of 0.1 between the LAT and the GBM
fluence and by extrapolating the MeV spectrum to the high energy
range \citep{Le(2009)}. Although \cite{Guetta(2011)} find some GRBs
that by extrapolating their MeV fluence to the GeV window should
have been detected by LAT, but were not. Therefore, putting upper
limits on the fluence ratio is crucial in order to test the nature
of the spectrum at high energies. This, in turn, will further
constrain the emitting mechanism. For example, in the case of pure
synchrotron, we might expect an extrapolation of the sub-MeV
spectrum, while for SSC we might see a rising spectral slope in the
sub-GeV. There has been one example of such a rising spectral slope
in the famous GRB 090902B \citep{Abdo b(2009)}, but  the overall
picture is still highly uncertain.

In an earlier study, \cite{Guetta(2011)} have used the non-detection
of GeV emission from the majority of Fermi GRBs, to put upper limits
on the GeV fluence of long GRBs. They find an upper limit on the
LAT/GBM fluence ratio  of less than unity for $60\%$ of GRBs. This
is consistent but not better than the EGRET-derived limits
\citep{Ando(2008)}. In this paper, we use a more subtle approach to
further constrain these limits. As the LAT detected GRBs are also
among the brightest GRBs in the GBM band \citep{Swenson(2010)}, we examine the brightest
group of GBM bursts with no LAT detection. Those are expected to be
the bursts with the  highest undetected GeV fluence. We obtain upper
limits on their GeV fluence and use them to gain more stringent
upper limits on the LAT/GBM fluence ratio. As the LAT emission is
extended we obtain limits for both the duration of the prompt MeV
emission, $T_{90}$, and for the extended duration of 600 sec. These
limits rule out SSC as a model for the prompt emission, and put
further constraints on many proposed emission mechanisms.

\section{The Sample}
\label{sec:sample} We consider all long ($T_{90}$ $\ge$ 2 seconds)
GRBs detected by GBM by February 2010. The GBM has a much wider
($\sim 8$sr) field of view than the LAT ($\sim 2.4$sr). After the
GBM detects a bright enough GRB, according to some selection criteria, Fermi
slews the LAT bore-sight towards
the GBM position. We consider only GBM-detected bursts within the LAT field of view.
To minimize the effect of slewing (towards the burst or due to the regular motion of the spacecraft)
we consider only bursts with small bore-sight angles.
It should be stressed that in the following analysis, we do not 
try to mimic the complicated slewing behaviour of the burst, but rather
we develop a method which largely overcomes the difficulties posed by this slewing.
As mentioned in \S \ref{sec:int}, those GRBs with a LAT detection, are
the brightest GRBs in the GBM range. Therefore we expect the
brightest GBM bursts with no GeV detection to be the brightest GeV
bursts undetected by LAT. These are also the bursts for which we can
obtain the tightest limits on the sub-MeV to Gev fluence ratio.

Taking the above considerations into account, our sample of bursts
detected only by the GBM (denoted as the "GBM sample") includes the
18 most fluent long GRBs (within the LAT viewing angle and with a
bore-sight angle of less than $60^o$ \footnote{ The
effective area decreases rapidly for larger bore-sight angles.
This selection criteria reduces the errors in the fluence
due to the varying bore-sight angle.}) that were detected by the GBM
and undetected by LAT. This corresponds to bursts with GBM fluence
over $8.3\times10^{-6}$ erg/cm$^2$. The seemingly random choice
of lower limit on the fluence is taken from the number of bursts we consider.
If we were to find any significant GeV signal, it would have made sense to change
this limit in accordance and look at additional bursts, but this was not the case.
For all bursts in the sample, we
take all the $>$100MeV photons (as seen by the LAT) whose positions
in the sky are up to $15^o$ from the GBM position. The GBM
position error is $\sim10^o$ and can be even less for bright bursts
\citep{Briggs(2009)}.  Therefore, the source of the LAT
fluence (if such a source exists) is very likely within the data
set.  \footnote{As a check to this statement, we found that we obtain similar results
when taking photons from up to $10^o$ from the GBM position only}
We only take photons with a bore-sight angle of less than
$60^o$ as the LAT effective area per photon decreases rapidly for
viewing angles larger than $60^o$.

We also consider a sample of LAT detected bursts (denoted the "LAT
sample") that consists of the 10 bursts with a reported detection by
the LAT team during the same period. A GRB is tagged as detected by
the LAT if the number of photons detected, $N_s$, exceeds 10 and if
it exceeds a $5\sigma$ fluctuation of the background
(\citealt{Band(2009)}, \citealt{Atwood(2009)}). These bursts
include: 080916C, 080825C, 090217,090323, 090328, 090626, 090902B,
090926A, 091003A, 091031. Note that so far the LAT team published
detailed fluence measurements only for four of those bursts:
080825C, 080916C, 090217, 090902B \citep{Abdo(2010)}.

\section{The model - overview}
Our goal is to obtain an upper limit  on the GeV fluence of each of
the bursts in the GBM sample. A complete approach involves
maximizing a general likelihood function, for which the free
parameters are: the burst's position, (x,y), the ratio, r, between
the burst's fluence and the total (burst+noise) fluence, and the
spectral parameters of both the burst $N \sim E^{-\alpha}$ and the
noise $N_{noise}\sim E^{-\beta}$. The likelihood function is:
\begin{equation}
 L'(x,y,r,\alpha,\beta) =  \prod_{i=1} A(E_i,\theta_i) P(x,y,x_i,y_i,E_i)[r N(E_i)+(1-r) N_{noise}(E_i)]\ ,
\end{equation}
where the summation is over all photons within $15^o$ of the
GBM position from t=0 to t=T (T can be $T_{90}$ or 600 sec depending
on the assumption on the origin of the GeV emission).  $x_i$, $y_i$
are the  right ascension and declination of the i'th photon in local
Cartesian coordinates.\footnote{The plane of right ascension and
declination (in which the photons locations are given) is not a flat
plane but a surface of a sphere. As we are only dealing with
locations within $15^0$ from the GBM location, we can (to a very
good approximation) use a locally flat coordinate system. This is
done by taking the projection of each position on the sphere to the
plane tangent to the sphere at the GBM position. These are the x,y
coordinates.} $P(x,y,x_i,y_i,E_i)$ is the probability that a
photon with energy $E_i$ from a burst at (x,y) will be detected at
$(x_i,y_i)$, namely it is the value of the LAT point spread function
for photon energy $E_i$ at the angle between (x,y) and $(x_i,y_i)$
\citep{Atwood(2009)}.

Unfortunately, this method turns out to depend  strongly  on the
unknown spectrum of the noise $N_{noise}$, and it cannot be used. We
turn to a different approach which is more stable. We split the
process into two phases. First we find the most probable positions
of the bursts using a simpler likelihood procedure. Then, using
these positions we calculate fluence upper limits.

\begin{figure}
\centerline{ \epsfig{file=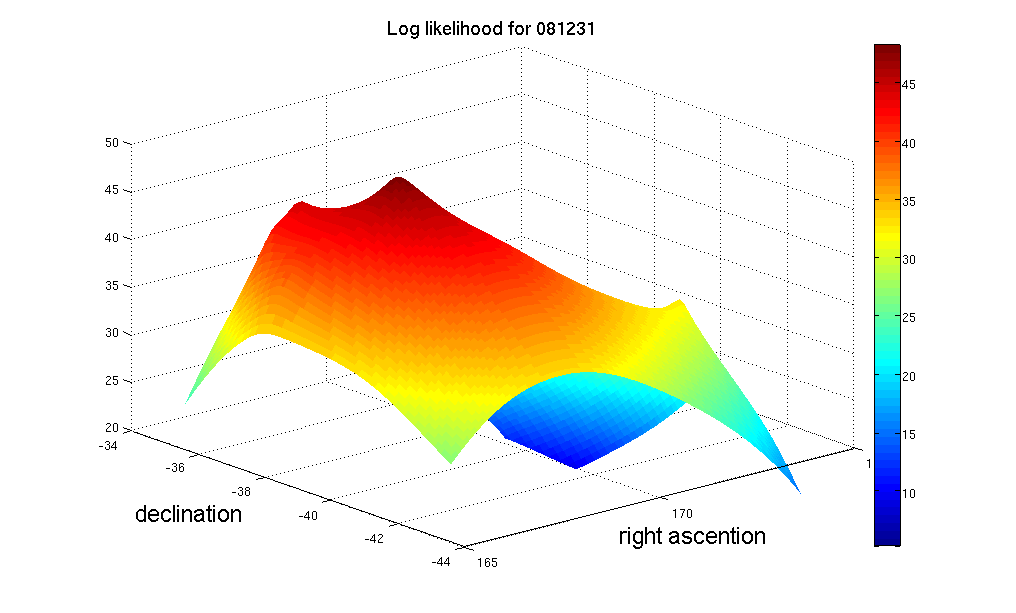, scale=0.5} } \caption
{\small The log of the likelihood function (Eq. \ref{eq position
likelihood}) for GRB 081231} \label{fig:like}
\end{figure}

\subsection{Bursts' positions}
\label{sec:detpos} We find the most probable position of a burst by
maximizing the likelihood, $L$, that it is located at (x,y):
\begin{equation}\label{eq position likelihood}
 L(x,y) =  \prod_{i=1} A(E_i,\theta_i) N(E_i) P(x,y,x_i,y_i,E_i)\ .
\end{equation}
Fig. \ref{fig:like} depicts an example of this likelihood function
for burst 081231. To calculate the likelihood function we need the
burst's spectrum, $N(E)$. We assume $N\propto E^{-2}$.  We allow for
variation in the slope of the spectrum and find that the dependence
of the position on this parameter is weak. A change of the spectral
index between -1 to -3, yields a change in the calculated position
of less than $0.1^o$. In addition, we verify in \S \ref{sec:LAT}
that the assumption $N\propto E^{-2}$ results in good agreement with
the positions found by the LAT team for LAT detected bursts in the
sample. Table \ref{tbl:pos} describes the positions of the bursts
and compares them to the GBM positions. Note that the GBM positions
have error bars of $10^o$ and that the most probable position that
we calculate may be dominated by the noise.
The errors in our locations are calculated according to the
decline of the log-likelihood function \citep{Press(1992)} and correspond
to 68$\%$ confidence levels.

\begin{table}\small
\begin{tabular}{ l p{2.5cm} p{2.2cm} p{2.3cm} p{2cm} p{2cm}}
\hline
Burst & Most probable LAT& Most probable LAT& error in& GBM  & GBM  \\
&  right ascension &  declination& location (deg)& right ascension &  declination \\
\hline
080906B & 184.3 & -2.3 & 0.4 &185.5 & -8.6 \\
080925  & 97.8  & 18.1 &0.9 &96.1 & 18.2 \\
081009  & 251.9 &17.8  &1.5 &251.1 & 17.1 \\
081207  & 132.1 & 63.6 & 3.7 & 119.2 & 66.9\\
081222  & 23.9  & -34.2&1.8 &22.7 & -34.1 \\
081231  & 218   & -38.1&1 &218 & -38.7 \\
090117C & 23.8   & -36.5& 2.6 &22.8 & -34.1 \\
090131  & 349.1 & 15.1& 1.6 &353 & 16.4 \\
090330  & 160.8 & -6  & 0.5 &159.2 & -7.7 \\
090516A & 139.6 &-15.2& 0.6&138.3 & -11.9 \\
090516B & 122.5 & -67.9& 0.3&122.2& -71.6 \\
090524  & 333.6 & -63.3& 2.3&329.5 & -67.4 \\
090528B & 312.3 & 35.9 & 2.8&312.2 & 32.7 \\
090711  & 141.8   & -59.4 & 0.3 &139.6 & -64.7 \\
090720  & 217.4 & -60 & 1.9 &203 & -54.8 \\
090829A & 330.3 &-33.8& 0.2 &329.2 & -34.2 \\
090922A & 13    & 72.1& 0.45 &17.1 & 74.3\\
091120  & 225.7 & -24.8&0.6 &224.8& -24.8 \\
\hline
\end{tabular}
 \caption{\small Positions of our "GBM sample" - 18 brightest
 GBM bursts in the LAT field of view that were not detected by the LAT.
 The method we use to calculate the most probable LAT position is described in the text}
\label{tbl:pos}
\end{table}

\subsection{Upper limits for the GBM sample}
\label{sec:uplim}

To obtain upper limits on (or estimate)  the GeV fluence we insert a
burst-like signal with a known fluence and a given position into the
actual data of the burst and mimic the process  of identifying the
burst's position using the maximum likelihood method (described in
\S \ref{sec:detpos}). If the new maximum is closer to the position
of the planted signal than to the position found in the original
data, the signal is considered to be identified. We repeat this
process many times, each time drawing a different realization of the
signal from a given distribution. The minimal fluence that the
artificial signal must have in order to be identified over the
actual burst signal (or background noise) by our method in more than
90$\%$  of the realizations is considered as an upper limit to the
burst's fluence at a confidence of 90$\%$.

We simulate a source-like signal, assuming a spectrum\footnote{We
have checked in \S \ref{sec:detpos} that the dependence of bursts'
positions on the spectrum are weak. In addition we verify here that
also the upper limits are not highly dependent on the spectrum.
As we raise the spectral slope, the signal consists of less energetic
photons. This makes the signal harder to detect and therefore raises
the limits we obtain. Changing the spectrum from $N\propto E^{-1}$ to
$N\propto E^{-3}$ yields an average increase of  $25\%$ in the upper
limits for 600 sec. In addition the various bursts
in the sample have very little spread around this average.}
$N\propto E^{-2}$. Given the spectrum, the fluence corresponds to an
average number of photons, $N_{avg}$, that composes the simulated
signal. We allow for random Poisson fluctuations of $\sqrt{N_{avg}}$
in  the number of photons and choose the energy of each photon
randomly with a probability defined by the spectrum. The right
ascension and declination of each photon are taken according to the
PSF and the angle to the detector is determined by the average
incidence angle of the actual incoming photons in the relevant time
interval at that position. The average bore-sight is taken by considering
all the photons arriving after t=0 and averaging over their effective areas.
The alternative, to give each photon a distinct bore-sight angle,
would have required us to make further assumptions regarding the temporal profile
of the signal, thus further complicating the analysis.
Notice, that even for the maximal slewing during a burst in the sample, which is about $50^o$,
the difference between the actual effective area and the average we use is less than a factor of two.
Hence, this is also the maximal error due to slewing in our estimated flux.
The consistency or our treatment of slewing is confirmed 
by the fact that we reproduce
the results for LAT bursts with published fluence (see \S \ref{sec:LAT}).

We center the simulated signal at a
random point between $5^o$ to $10^o$ away from the position
determined from the original data. This ensures that there is no
confusion with the data of the most probable source.

We plant the photons of the simulated signal onto the original data
from t=0 to t=T (where T is either $T_{90}$ or 600 sec) and again
apply the maximum likelihood method.
If the new maximum is closer to the  position of the planted signal
than to the position found in the original data, the signal is
considered as detected.
We repeat this process a 1000 times for
different fluxes and we note the value
 of $S_{90\%}$ for which the  detection probability of the inserted signal reaches 90$\%$ .

\begin{center}
\begin{table}\small
\begin{tabular}{|l||p{2.5cm}||p{2.4cm}||p{2.6cm}||p{2.4cm}||p{2.4cm}|}
\hline
Burst & $90\%$ upper limit at $T_{90}$ (erg/cm$^2$) & LAT/GBM fluence upper limit at $T_{90}$\footnotemark[1]& Band extrapolated fluence at $T_{90}$(erg/cm$^2$)
& $90\%$ upper limit at 600 sec (erg/cm$^2$) & LAT/GBM fluence upper limit at 600  sec\\
\hline
080906B & $9.27\times10^{-7}$   &0.09 & $8.44\times10^{-6}$&  $4.51\times10^{-6}$ &0.41 \\
080925  & $2.03\times10^{-6}$   &0.15  & $2.24\times10^{-6}$& $2.03\times10^{-6}$   &0.69\\
081009  & $2.12\times10^{-6}$         &0.26& $1.16\times10^{-13}$& $1.03\times10^{-5}$  &1.24 \\
081207  & $2.83\times10^{-6}$   &0.03 & $1.61\times10^{-5}$& $8.09\times10^{-6}$    &0.08 \\
081222  & $2.47\times10^{-6}$   &0.18 & $9.69\times10^{-6}$& $5.1\times10^{-6}$  &0.38 \\
081231  & $7.6\times10^{-7}$    &0.06 & $1.58\times10^{-5}$& $8.02\times10^{-6}$    &0.67 \\
090117C & $1.27\times10^{-6}$    &0.12 & $8.91\times10^{-6}$& $5.74\times10^{-6}$    &0.52 \\
090131  & $2.07\times10^{-6}$   &0.1 & $2.67\times10^{-6}$& $7.9\times10^{-6}$  &0.35 \\
090330  & $1.09\times10^{-6}$   &0.1 & $1.03\times10^{-7}$& $5.71\times10^{-6}$    &0.5\\
090516A & $4.28\times10^{-6}$        &0.19 & $8.26\times10^{-6}$& $4.77\times10^{-6}$   &0.21 \\
090516B & $5.31\times10^{-6}$   &0.18 & N/A\footnotemark[2]& $8.57\times10^{-6}$    &0.29 \\
090524  & $2.54\times10^{-6}$   &0.13 & $1.82\times10^{-6}$& $7\times10^{-6}$   &0.38 \\
090528B & $4.1\times10^{-6}$    &0.09 & $6.06\times10^{-6}$& $1.24\times10^{-5}$    &0.27\\
090711  & $1.7\times10^{-6}$      &0.15 & N/A& $1.7\times10^{-6}$     &0.15 \\
090720  & $3.63\times10^{-6}$   &0.34 & $1.71\times10^{-6}$& $1.28\times10^{-5}$     &1.2\\
090829A & $2.67\times10^{-6}$   &0.03& $6.5\times10^{-5}$& $4.1\times10^{-6}$  &0.04\\
090922A & $3.23\times10^{-6}$   &0.28 & $1.73\times10^{-6}$& $1.04\times10^{-5}$    &0.91\\
091120  & $2.46\times10^{-6}$          &0.08& $1.51\times10^{-8}$& $8.4\times10^{-6}$   &0.28 \\
\hline
\end{tabular}
 \caption{\small Upper limits at $90\%$ confidence level on the GeV fluence of the bursts in the GBM
 sample, the LAT/GBM fluence ratio and LAT fluence assuming an
 extrapolation of the GBM Band spectrum. The GBM fluence and spectrum
 used for extrapolation is taken from Zhang et al., 2010 }
\footnotemark[1]{The GBM Fluence was taken in the range of
8-1000keV}

\footnotemark[2]{Band parameters are not available}
\label{tbl:flu}
\end{table}
\end{center}

\section{LAT detected bursts}
\label{sec:LAT} As a test for the overall method we apply it to the
sample of 10 LAT detected bursts. First, we compare the positions of
the LAT bursts we find using the maximum likelihood method, to those
reported by the LAT team. A good agreement (a deviation of about
$0.5^o$) is found within the errors of the method with most of the positions
reported by the LAT team (see Table \ref{tbl:pos2}).
Notice that 080825C has a large bore-sight angle ($\theta=60^o$), resulting
in a large fraction of photons with extremely small effective areas.
This leads to fewer photons (relative to bursts with small
bore-sight angles and with a similar flux) and at the same time to an
increase in the photon's weighted flux ($E/A(E,\theta)$). 
The result is a large increase in the statistical error.
Indeed the deviation from the LAT position is large in this
case ($2.1^o$) but within the estimated error bars.
The overall agreement of the  positions found with those published by
LAT team is an additional confirmation that the uncertainty in the
assumed spectrum of the source isn't crucial in determining the
positions, and the results based on the assumption $N\propto E^{-2}$
are valid.

\begin{table}\small
\begin{tabular}{ l p{2cm} p{1.6cm} p{1.4cm} p{1.1cm} p{1.7cm} p{2.5cm} p{2cm}}
\hline
Burst & calculated right ascension & calculated declination& LAT right ascension & LAT declination & LAT Bore-sight angle (deg) &
difference between LAT and calculated locations (deg) & error in location estimate (deg)\\
\hline
080916C & 117.5 & -56.5 &119.8 & -56.6 & 48 & 2.3 & 2.5\\
080825C & 233   & -2.8  &233.9 & -4.7 & 60 & 2.1 & 3.6\\
090217  & 205.5 &-8.4   &204.9 & -8.4 & 34 & 0.6 & 0.9\\
090323  & 190.5 & 17     & 190.7 & 17 & -1 & 0.2\footnotemark[1] & 0.5\\
090328  & 91    & -42   &90.9 & -42 & -1 & 0.1\footnotemark[1] & 0.25\\
090626  & 171.1 & -33.3 &170 & -33.5 & 15 & 1.1 & 1.5\\
090902B & 265.7  & 27.3 &265 & 27.3 & 52 & 0.7 & 0.6\\
090926A & 353.8 & -66.3 &353.6 & -66.3 & 52 & 0.2 & 0.3\\
091003A & 253.2 & 36.5  &251.4 & 36.6 & 13 & 1.6 & 0.8\\
091031  & 71.7  &-57.6  &71.7 & -57.5 & 22 &0.1 & 0.2\\
\hline
\end{tabular}
 \caption{\small Positions of LAT-detected GRBs found by our
 calculation compared to those reported by the LAT team (\citealt{Tajima(2008)}, \citealt{Bouvier(2008)}, \citealt{Ohno a(2009)}, \citealt{Ohno b(2009)},
\citealt{McEnery a(2009)}, \citealt{Piron(2009)}, \citealt{de Palma a(2009)}, \citealt{Uehara(2009)}, \citealt{McEnery b(2009)}, \citealt{de Palma b(2009)})}
\footnotemark[1]{No bore-sight angle available}
\label{tbl:pos2}
\end{table}

We obtain upper limits and fluence estimates for all LAT bursts. We use the ``planted
signal" method for the upper limits at a confidence level of $50\%$
instead of $90\%$. A signal that in $50\%$ of the cases is detected
over the original signal must be comparable to the main peak in the
data. If one assumes that this peak originates from the burst itself
rather than from the noise, the fluence of such a signal should be
an estimate to the original burst's fluence.
Again 080825C is an exception.

\begin{figure}
\centerline{ \epsfig{file=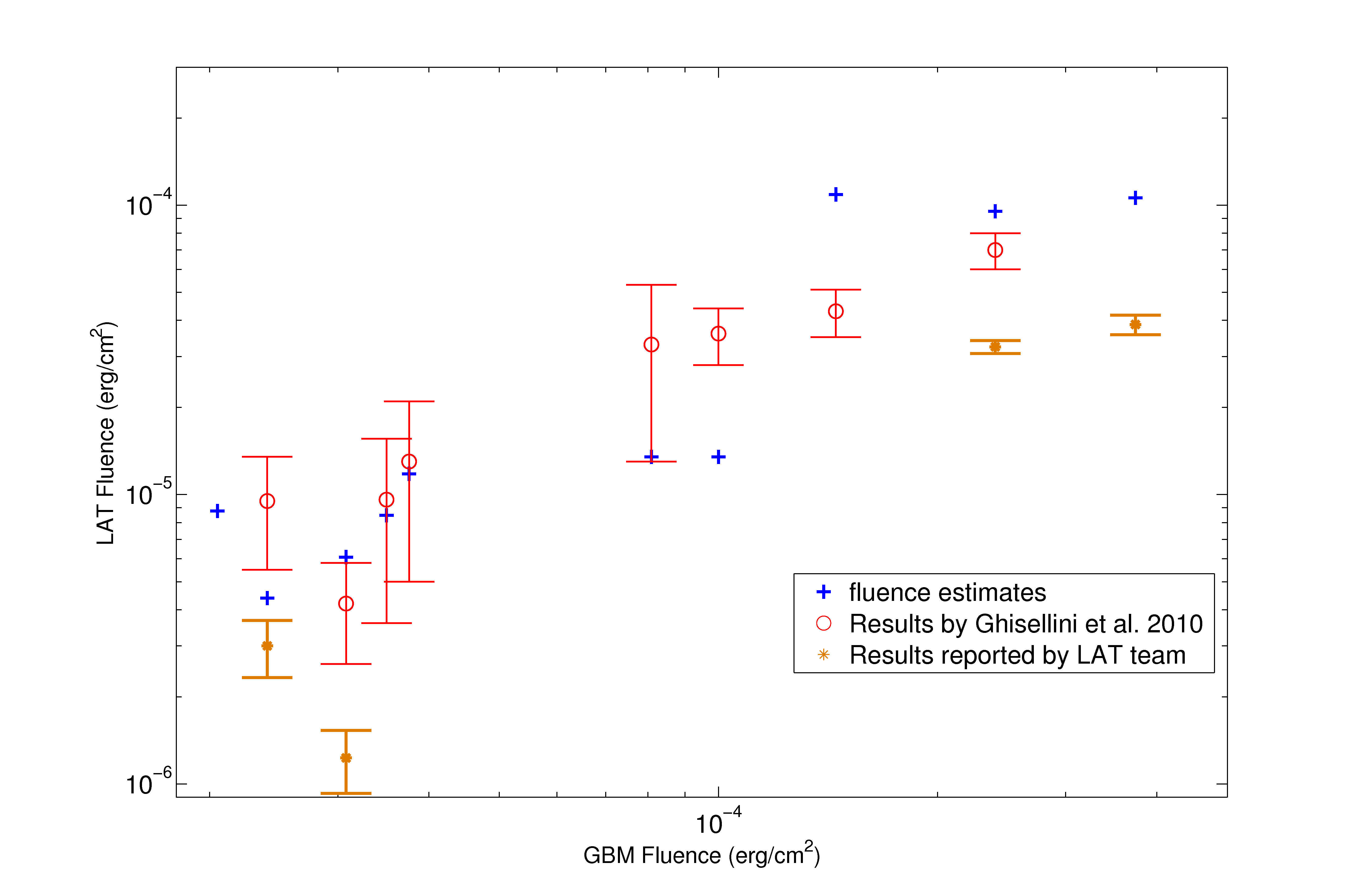, scale=0.5} } \caption
{\small LAT detected bursts at 600 sec: Shown are the fluence
estimates. These are
compared to results by Ghsellini et al. 2010 and to results reported
by the LAT team.} \label{fig:latdet600}
\end{figure}

To date, the LAT team published fluence measurements to only four of
these bursts \citep{Abdo(2010)}. We compare the upper limits and
fluence estimates to these bursts in Table \ref{tbl:comp}. Our upper
limits are consistent with the fluence estimates and with the LAT
results. It is noticeable that our
estimated fluences tend to overestimate the LAT reported fluence by
a factor of $\sim 3$. This is possibly due to extra noise cuts
applied by the LAT team using a careful treatment of each photon
separately. When considering only upper limits to individual bursts
this doesn't cause a problem as it simply makes the upper limits more
conservative. We also compare our fluence estimates to those of
\cite{Ghisellini(2010)} and find comparable values (figure \ref{fig:latdet600}).
Note that \cite{Ghisellini(2010)} also overestimate the fluences
reported by the LAT team by a factor of 2-4.

\begin{figure}
\centerline{ \epsfig{file=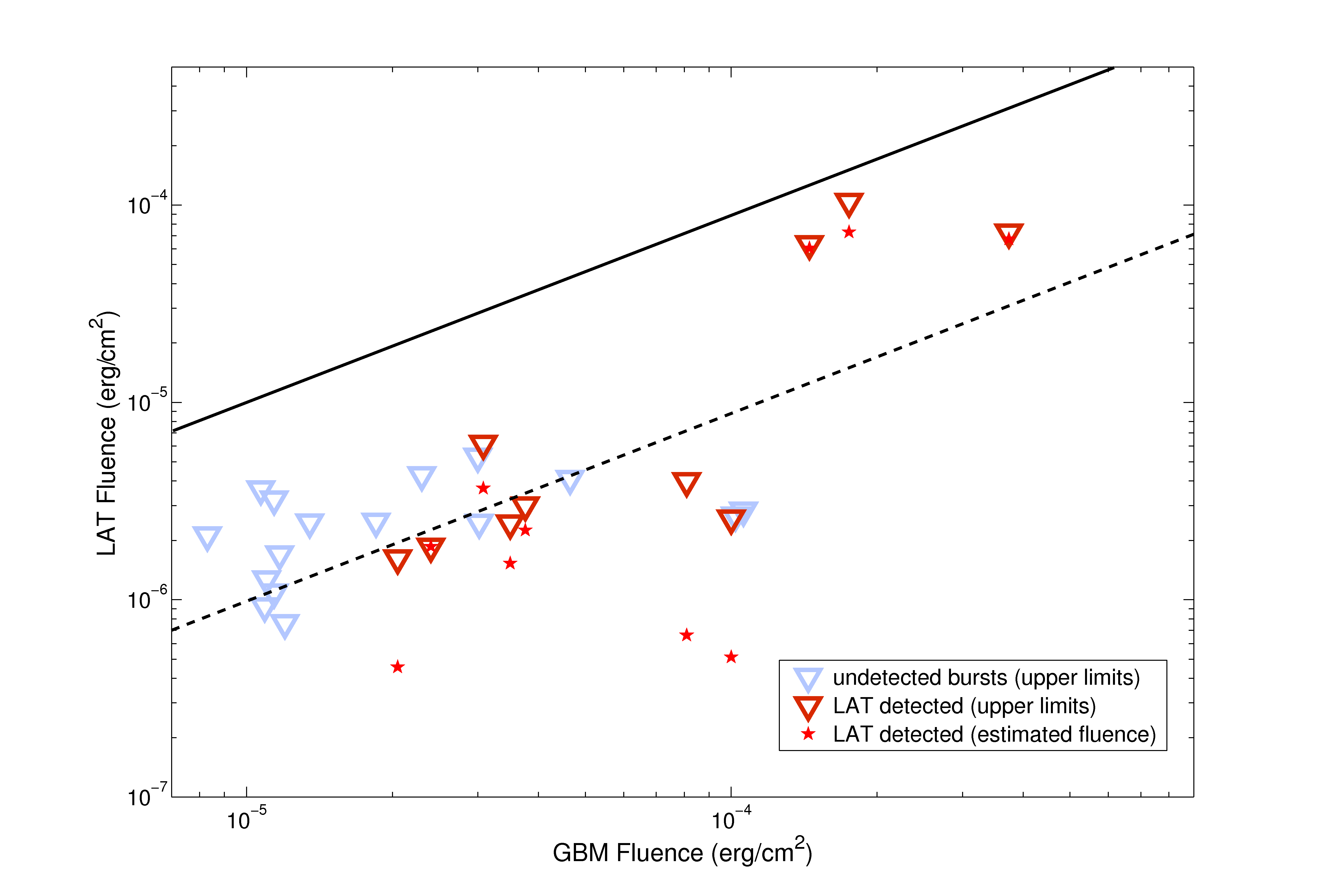, scale=0.5} } \caption {\small
The LAT fluence vs. the GBM fluence  for two types of bursts: LAT
undetected (the GBM sample) and LAT detected (the LAT sample)
bursts, at $T_{90}$ duration. For the GBM bursts we provide upper
limits at a $90\%$ confidence level, while for LAT bursts we provide
both our upper limits and estimates of the fluence. The solid line
marks equal fluence in the LAT and GBM bands, and the dotted line
marks a LAT/GBM fluence ratio of 0.1.} \label{fig:latgbmt90}
\end{figure}

\begin{figure}
\centerline{ \epsfig{file=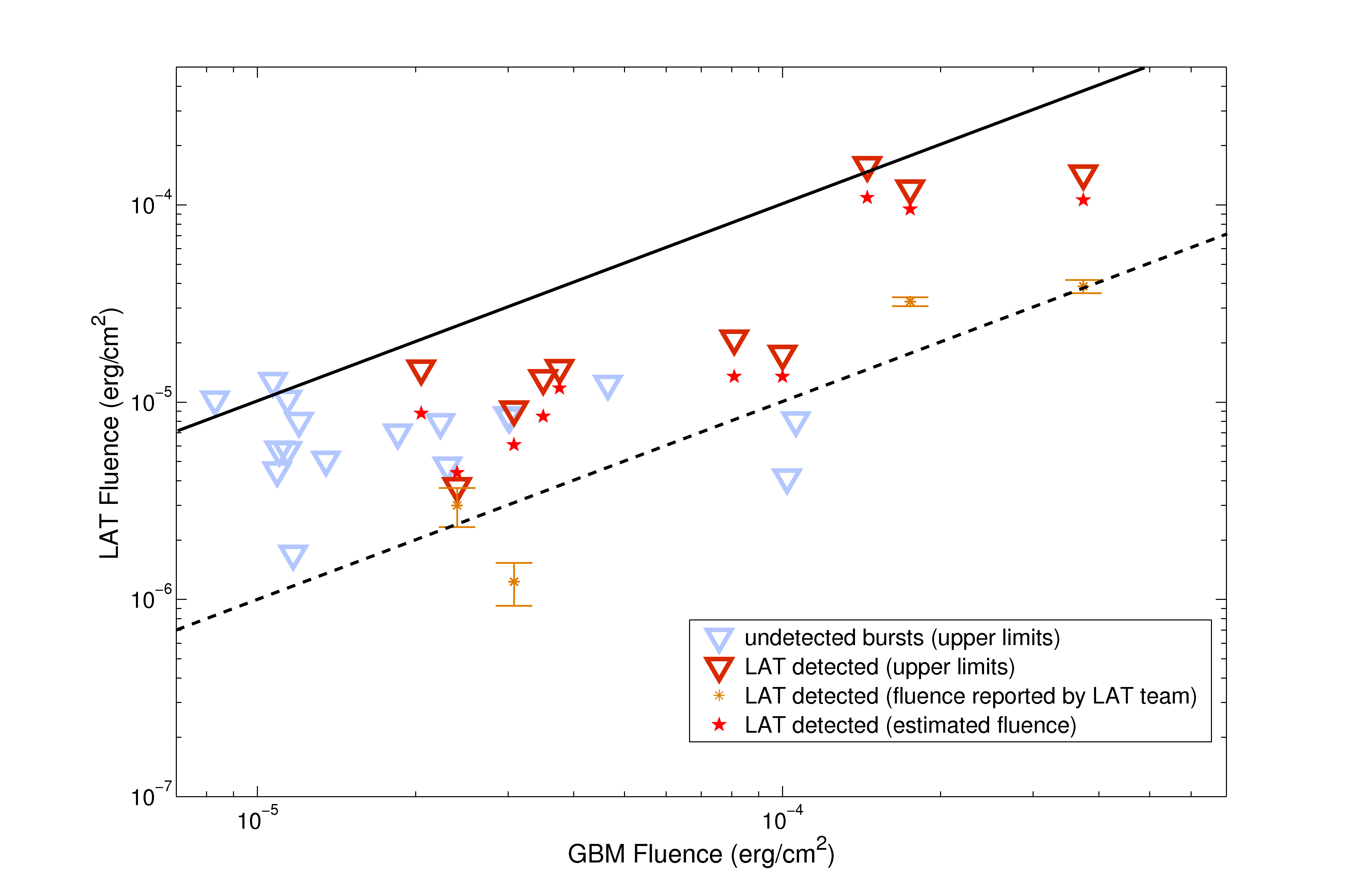, scale=0.5} } \caption {\small
The LAT fluence vs. the GBM fluence  for two types of bursts: LAT
undetected (the GBM sample) and LAT detected (the LAT sample)
bursts, during the first 600 sec after the trigger. For the GBM
bursts we provide upper limits at a $90\%$ confidence level, while
for LAT bursts we provide both upper limits and actual estimates of
the fluence. Also depicted are 4 fluence estimates reported by the
LAT team. The solid line marks equal fluence in the LAT and GBM
bands, and the dotted line marks a LAT/GBM fluence ratio of 0.1.}
\label{fig:latgbm600}
\end{figure}

Figs. \ref{fig:latgbmt90} \& \ref{fig:latgbm600}  depict comparisons
of the  LAT/GBM fluence ratios of LAT bursts with  the
upper limits on the LAT/GBM fluence ratio of LAT non-detected bursts
for  $T=T_{90}$ and T=600 sec. The average {LAT}/{GBM} fluence ratio
for LAT detected bursts is  0.09 (0.2) for $T_{90}$ (T=600 sec).
These limits are somewhat lower than the corresponding upper limits
on these ratios derived for the GBM bursts.

\begin{table}\small
\begin{tabular}{ l p{2.8cm} p{2.8cm} p{3.6cm} }
\hline
Burst & $90\%$ Upper limit (erg/cm$^2$) & $50\%$ detection estimate (erg/cm$^2$)  & Fluence reported by the LAT team (erg/cm$^2$)\\
\hline
080916C &  $1.2\times10^{-4}$& $9.5\times10^{-5}$  & $3.2\times10^{-5}$  \\
080825C &  $7.3\times10^{-6}$& $4.4\times10^{-6}$ & $3\times10^{-6}$   \\
090217  &  $9.2\times10^{-6}$& $6.1\times10^{-6}$ & $1.2\times10^{-6}$ \\
090902B &  $1.4\times10^{-4}$& $1\times10^{-4}$  &$3.9\times10^{-5}$    \\
\hline
\end{tabular}
 \caption{\small A comparison of the 600 sec fluences and upper limits that we
 calculate for LAT detected bursts, to those reported by the LAT team. The LAT reports are
 for the total burst fluence (not limited by time after the trigger),
 but these are dominated by the fluence during the first 600 sec.}
  \label{tbl:comp}
\end{table}

\section{Results}
\label{sec:RES} The results for the duration of $T_{90}$ are
summarized in table \ref{tbl:flu}. For the GBM bursts we find upper
limits with an average fluence of
$S_{90\%}(T_{90})=2.4\times10^{-6}$ erg/cm$^2$, corresponding to an
average upper limit of 0.13 on the LAT/GBM fluence ratio.
Notice (Fig. \ref{fig:latgbmt90}) that these limits are almost
uniform for all GBM bursts and do not seem to depend on the GBM
fluence of the bursts. This means that the upper limits derived here
are mainly representative of the LAT detection limit (with our
method) and do not show any evidence for actual GeV signals in the
GBM sample.
The upper limits at 600 sec has an
average fluence of $S_{90\%}(600 s)=7.3\times10^{-6}$ erg/cm$^2$,
corresponding to an upper limit of 0.45 on the  LAT/GBM fluence
ratio. As for the case of $T_{90}$, Fig. \ref{fig:latgbm600} shows
almost uniform limits for different GBM fluences. Therefore these
limits too, correspond to the LAT detection limit within our method.
The fact that the limits here are higher than the $T_{90}$ limits is
merely an artifact of the longer timescales and therefore higher
noise level. These results are summarized in table \ref{tbl:flu}.

\begin{figure}
\centerline{ \epsfig{file=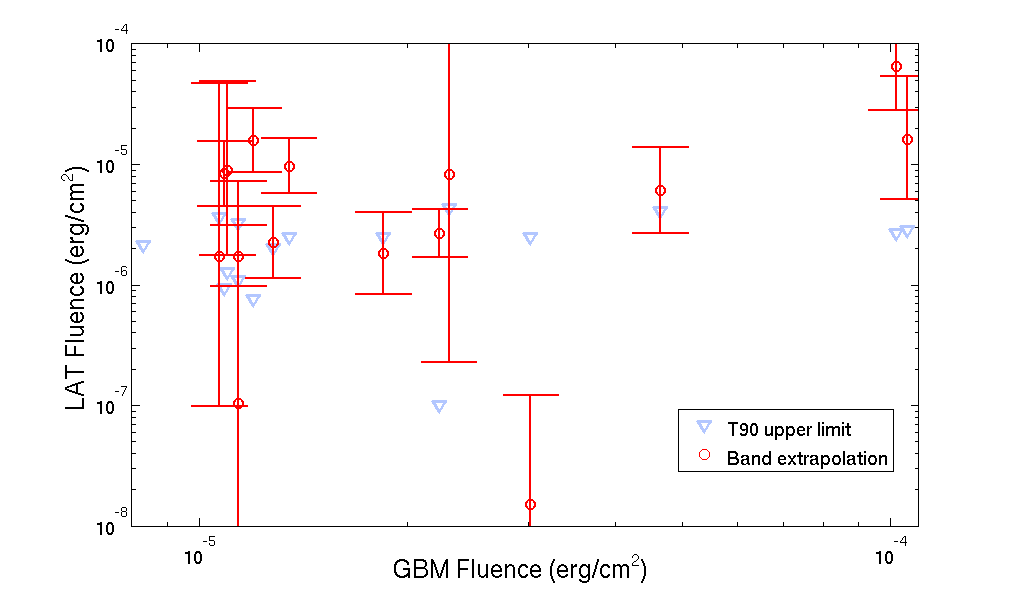, scale=0.4} } \caption {\small
The 100MeV - 10GeV fluence obtained by extrapolation of the GBM Band spectrum to GeV for
the GBM sample, compared with the LAT upper limits on the GeV
fluence. For two thirds of the sample, the extrapolation of the Band
function is consistent with the upper limits. The extrapolated
fluence of the remaining five bursts is significantly larger than
their upper limits. These bursts  suggest a decline in the spectral
slope between the MeV and the GeV. 090516B and 090711 have no Band
parameters reported in the GCNs (GCN data: \citealt{Bissaldi a(2008)}, \citealt{Goldstein a(2008)},
\citealt{von Kienlin(2008)}, \citealt{Briggs(2008)}, \citealt{Bissaldi b(2008)}, \citealt{Goldstein b(2008)},
\citealt{Connaughton(2009)}, \citealt{Goldstein(2009)}, \citealt{von Kienlin a(2009)}, \citealt{McBreen a(2009)},
\citealt{McBreen b(2009)}, \citealt{von Kienlin b(2009)}, \citealt{von Kienlin c(2009)}, \citealt{Rau a(2009)},
\citealt{Wilson(2009)}, \citealt{Rau b(2009)}, \citealt{Gruber(2009)})}
\label{fig:extrap}
\end{figure}

The overall number of bursts detected by LAT is consistent with an
extrapolation of the Band function fitted at the MeV range
\citep{Le(2009)} to the GeV range. It is interesting to check
weather this extrapolation is consistent with the upper limits we
obtain in \S \ref{sec:uplim}. For all bursts in the sample, we take
the Band parameters (and corresponding errors) \citep{Band(1993)}:
$E_{peak}$ (the peak in $\nu F_{\nu}$), $\alpha$ (the lower energy
spectral slope), $\beta$ (the higher energy spectral slope) as
reported in the GCNs. In addition we also take
the bursts' $T_{90}$ and the overall fluence in the GBM band. We
extrapolate the MeV emission of GBM bursts using the Band function
to the LAT energy range [100 MeV - 30 GeV]. We
calculate the overall LAT fluence expected for each burst in the
sample assuming no break in the spectral index between the GBM and
the  LAT windows.
As we do not have the covariance matrices of the spectral fits,
and we don't know the degree of correlation between the various errors,
we assume the "worst case scenario".
This is to say we take those correlations that maximize the overall error
in the fluence estimate. Specifically, this means that when looking for a lower
limit on the extrapolation we take the lower values of $N_0$ and $E_p$ and the higher value of $\beta$.
Clearly, changing $\alpha$ is also not relevant for the sake of 
these limits. This is what we refer to as $1\sigma$, it is an
overestimate of the actual value. As can be seen in table \ref{tbl:flu} and in Fig.
\ref{fig:extrap} most expected fluences (9 out of 16) are compatible
with the upper limits. In two bursts (e.g. 081009, 091120) the Band
extrapolations are a few orders of magnitudes below the upper
limits. This is not necessarily significant as these are only upper
limits and the actual LAT fluence is unknown and can be low. The
rest (5 out of 16), almost a third of the bursts, have Band
extrapolations significantly above the upper limits on their
$T_{90}$ fluence. In most of these bursts the Band
extrapolation is 2-3$\sigma$ above our LAT 90$\%$ upper limit.
Therefore, there is no single burst in which this excess is measured
without doubt. However the fact that such a non-negligible access is
observed in a large fraction of the sample is notable, and it
suggests an evidence for a decline in the slope of the spectrum at
high energies of some bursts.

\begin{figure}
\centerline{
\epsfig{file=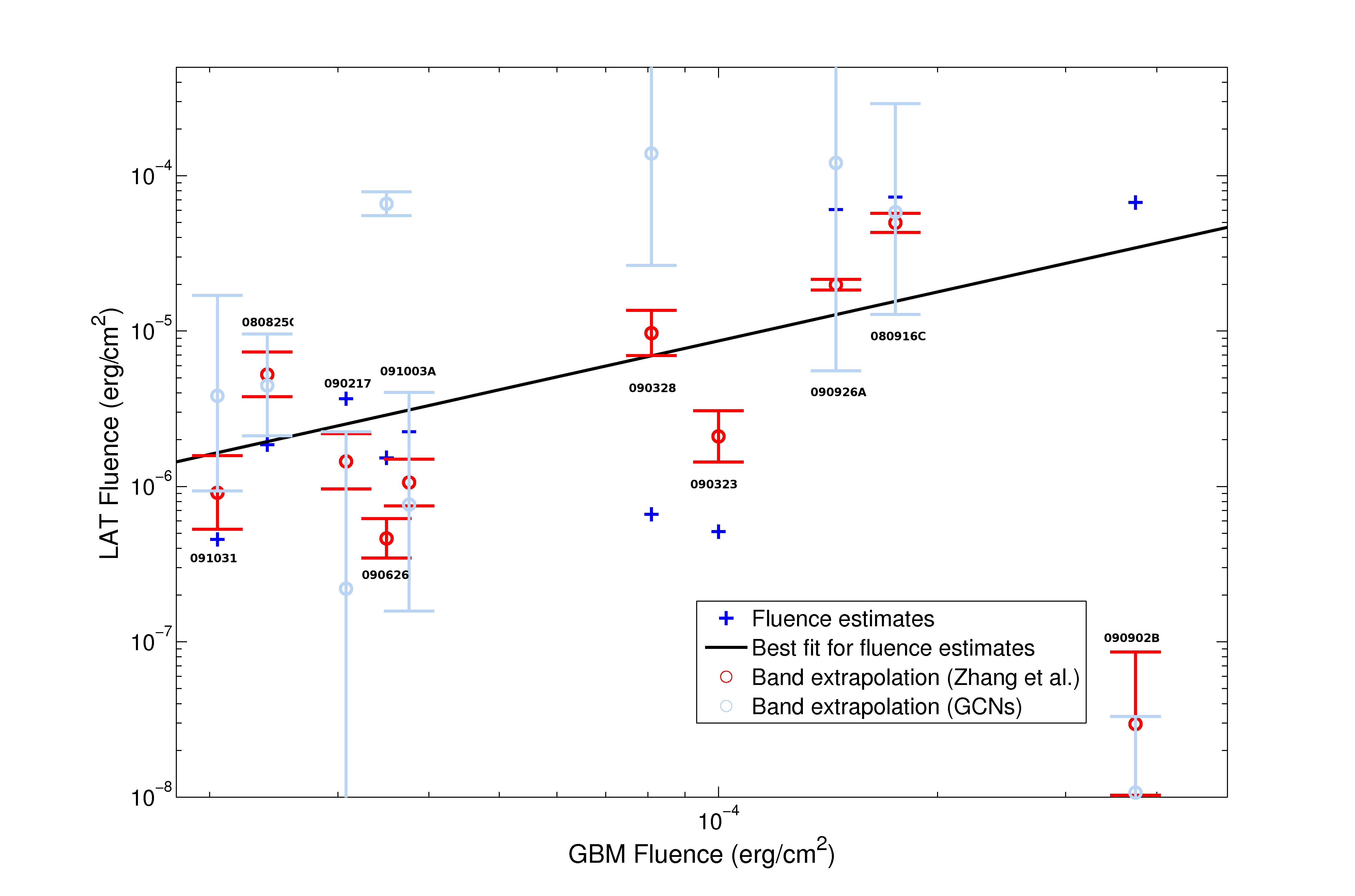, scale=0.5}
}
\caption {\small  LAT detected bursts at $T_{90}$. Shown are the fluence estimates
and the Band function extrapolations to the GeV range (using Band parameters either from Zhang et al. 2010 or from the GCNs).
The solid line marks the best fit for the fluence estimates, it's slope is close to 1.
This figure shows a positive correlation between LAT and GBM fluence of LAT bursts.
GRB 090902B has a low extrapolated fluence of $\sim 10^{-8}$ which falls below the figure.
GRB 090323 has no reported Band parameters in the GCNs. Three bursts, whose Band extrapolations are  larger than the observed LAT fluences,  show evidence for  a break in the high energy spectral slope
(GCN parameters: \citealt{van der Horst a(2008)}, \citealt{van der Horst b(2008)}, \citealt{von Kienlin d(2009)}, \citealt{Rau c(2009)}, \citealt{von Kienlin e(2009)}
, \citealt{Bissaldi a(2009)}, \citealt{Bissaldi b(2009)}, \citealt{Rau d(2009)}, {Golenetskii(2009)}).}
\label{fig:latdett90}
\end{figure}

A similar pattern is observed when we extrapolate MeV emission using
the Band function reported by \cite{Zhang(2010)} for the LAT
detected bursts. The comparison of the Band extrapolated fluence and
our estimates of the actual $T_{90}$ LAT fluence is depicted in Fig.
\ref{fig:latdett90}, where we also present Band extrapolations based
on fits reported in GCNs, that in almost all cases\footnote{The two
sets of results are generally simillar except for the case of GRB
090626 where the calculation according to the GCN parameters yields
a fluence two orders of magnitude lower then that obtained with the
\cite{Zhang(2010)} parameters. Notice that for GRB 090323 there are
no Band parameters in the GCNs.} agree with those of
\cite{Zhang(2010)}. For the majority of the bursts (080916C, 090217,
090626, 090926A, 091003A, 091031) the Band extrapolated fluence is
of order of the LAT estimated fluence. These bursts are consistent
with a single spectral component ranging from MeV to GeV
\citep{Zhang(2010)}. In GRB 090902B, the extrapolated fluence is
three orders of magnitude lower than its estimated fluence,
indicating on an additional source of high energy emission
\citep{Abdo b(2009)}. In the last three bursts (080825C, 090323,
090328) the extrapolated fluence to the GeV range is  about an order
of magnitude higher ($\sim 3 \sigma$) than the estimated GeV fluence.
These cases provide again, evidence for a steepening of the high energy
spectral slope between the MeV and the GeV bands. Note that GRB
080825C has a large bore-sight angle and therefore our estimates are
less reliable for this burst.  However, the Band extrapolation for this burst is
also above the fluence reported by the LAT team for the entire LAT
fluence (not only confined to $T_{90}$) from this GRB.

Even though the break in the spectrum is not confirmed for any
of the bursts individually, we can attempt to estimate the Lorentz
factors for those GRBs where we can see a possible steepening in the
high energy spectral slope. This is done by assuming that the
steepening originates from the increase in the optical depth with
photon energy until it reaches order unity at the energy where a
break is seen in the spectrum \citep{LithwickSari01,Abdo a(2009)}.
We assume that the energy of the break is of the order of 1GeV for
these bursts (an increase by a factor of 10 in the break energy,
will only change the estimate by a factor $\sim1.5$). This yields
estimated Lorentz factors of: $\Gamma=170$ for 090323 and
$\Gamma=190$ for 090328. GRB 080825C has no known redshift, and
therefore we can not directly estimate it's Lorentz factor. For a
generic z=1 we obtain $\Gamma=220$. These results are within the
``generic range" expected for the Lorentz factor in GRBs.

The results rule out any model in which there is a strong GeV
component in the prompt mission. In particular they limit strongly
SSC models for the prompt emission as those will suggest a second
SSC component at the GeV (\citealt{Ando(2008)}, \citealt{Piran
Sari(2009)}). They pose a strong limit on the conditions within the
emitting regions showing no IC GeV Peak.

\section{Conclusions}
\label{sec:Con} GeV emission from GRBs is as of yet relatively
unexplored observationally. Up to February 2010, only 10 bursts were
detected by LAT in the GeV range. Already in this early
observational stage, though, there is much information that can be
extracted by detailed analysis. This, in turn, provides us with
independent tests to the various emitting mechanisms.

We have analyze the group of bursts expected to have the highest
(undetected) GeV component. Those are the 18 most luminous GBM
bursts with no LAT detection. For these bursts, we obtain upper
limits on the GeV fluence. For the whole sample we obtain an average
upper limit to the fluence ratio of 0.13 during the prompt phase
($T_{90}$) and an average upper limit to the ratio of 0.45 for 600
sec. These ratios strongly constrain various emission models and in
particular rule out SSC models for either the  MeV emission or the
GeV component in the prompt emission. In both cases, a significant
LAT component is expected.

The fluence estimates for LAT bursts, lead to somewhat lower ratios
for the LAT/GBM fluence ratio. These are $0.09 \pm 0.03$ and $0.2 \pm 0.09$ for the
durations of $T_{90}$ and 600 sec respectively (Figs. \ref{fig:latdett90}$\&$ \ref{fig:latdet600}). In addition, the LAT
bursts show a correlation between their LAT and GBM fluences, namely
- the stronger a burst is in the GBM band, the stronger it is in the
LAT band. Considering that  for the LAT undetected bursts there are
only  upper limits, These results are consistent with the LAT-GBM
fluence correlation which is seen in the LAT-detected bursts.
Namely, the LAT bursts don't have to be drawn from a different
population than the GBM bursts.

For the majority of the GBM and the LAT bursts the GeV fluence is compatible with the Band extrapolation of the MeV emission. 
However, out of the LAT bursts, in three cases the Band extrapolation of the MeV emission is higher than the observed fluence.
Similarly the Band extrapolation is higher than the LAT upper limits in 5 out of 16 GBM bursts.
Both results are consistent and suggest that in some bursts we observe
a decline in the spectral high energy  slope between the MeV and the GeV.
This may be the first indication for the long sought after pair opacity break in the high energy spectrum.
If so it can enable us to estimate corresponding values of the bulk Lorentz factors which are around a few hundred.

This research was supported by an ERC
advanced research grant, by the Israeli center for Excellent for High Energy AstroPhysics, by the Israel Science Foundation (grant No.
174/08) and by an IRG grant.

\end{document}